\documentclass[10pt,runningheads]{llncs}
\usepackage[margin=1.0in]{geometry}
\usepackage[T1]{fontenc}
\usepackage{cite}
\usepackage{amsmath,amssymb,amsfonts}
\usepackage{graphicx}
\usepackage{textcomp}
\usepackage{xcolor}
\usepackage{amssymb}
\usepackage{standalone}
\usepackage{algorithm}
\usepackage{algpseudocode}
\usepackage{amsmath}
\usepackage{etoolbox}
\usepackage{float}

\usepackage{comment}
\usepackage{microtype}
\usepackage{forest}
\usepackage{pgf-umlsd}
\usetikzlibrary{positioning, shadows, decorations.pathmorphing, fit, shapes.geometric, arrows.meta}
\usepackage{subcaption}
\usetikzlibrary{backgrounds}

\pgfdeclarelayer{background}
\pgfdeclarelayer{L1}
\pgfdeclarelayer{L2}
\pgfdeclarelayer{L3}
\pgfdeclarelayer{foreground}

\pgfsetlayers{background, L3, L2, L1, main, foreground} 

\usepackage{tikz}
\def\BibTeX{{\rm B\kern-.05em{\sc i\kern-.025em b}\kern-.08em
    T\kern-.1667em\lower.7ex\hbox{E}\kern-.125emX}}

\usepackage[T1]{fontenc}
\usepackage{graphicx}
\begin{document}
\pagestyle{empty}
\title{RADAR-Q: Resource-Aware Distributed Asynchronous Routing for Entanglement Distribution in Multi-Tenant Quantum Networks}
\titlerunning{RADAR-Q}
%
\author{Chenliang Tian\inst{*1} \and
Zebo Yang\inst{2} \and
Raj Jain\inst{1} \and Ramana Kompella\inst{3} \and Reza Nejabati\inst{3} \and Eneet Kaur\inst{3} \and Aiman Erbad\inst{4} \and Mohamed Abdallah\inst{5} \and Mounir Hamdi\inst{5}}
\authorrunning{C. Tian et al.}
%
\institute{
Department of Computer Science and Engineering, Washington University in St. Louis, USA \and
Department of Electrical Engineering and Computer Science, Florida Atlantic University, USA \and
Quantum Lab, Cisco, USA \and
Department of Computer Science and Engineering, Qatar University, Qatar \and
Department of Science and Engineering, Hamad Bin Khalifa University, Qatar \\
\email{chenliang.t@wustl.edu, yangz@fau.edu, jain@wustl.edu, rkompell@cisco.com, rnejabat@cisco.com, ekaur@cisco.com, aerbad@qu.edu.qa, moabdallah@hbku.edu.qa, mhamdi@hbku.edu.qa} 
}
\maketitle              

\let\thefootnote\relax\footnotetext{Submitted to the Fifth International Conference on Innovations in Computing Research (ICR'26), August 2026, Berlin, Germany.}
\begin{abstract}
Scalable quantum networks must support concurrent entanglement requests from multiple users, yet existing routing protocols fail when users compete for shared repeater resources and waste fragile quantum states that decay rapidly and cannot be buffered like classical data. This paper presents RADAR-Q, a resource-aware decentralized routing protocol that embeds real-time resource contention directly into path selection. Unlike prior designs that either require global coordination or route all traffic through a central anchor, RADAR-Q makes intelligent local decisions by balancing three factors: (1) path length and link fidelity, (2) instantaneous availability of quantum memory at each node, and (3) the number of intermediate Bell-State Measurement (BSM) operations needed to connect a source--destination pair. By identifying the Nearest Common Ancestor (NCA) within a DODAG routing hierarchy, RADAR-Q localizes entanglement swapping close to the communicating users---avoiding unnecessary detours through the network center and reducing both the BSM chain length and qubit exposure to decoherence.

We evaluate RADAR-Q on grid and random topologies---representing regular and irregular network fabrics, respectively---against state-of-the-art synchronous and root-centric asynchronous baselines. Results demonstrate that RADAR-Q achieves $2.5\times$ and $7.6\times$ higher aggregate throughput than synchronized and root-centric asynchronous designs, respectively. While baseline protocols suffer catastrophic fidelity collapse below the 0.5 distillation threshold under high load, RADAR-Q consistently maintains end-to-end fidelity above 0.76---ensuring every generated pair remains physically usable for downstream quantum applications. Furthermore, RADAR-Q exhibits near-perfect fairness (Jain's Fairness Index 96--98\%) and retains over 50\% of its ideal throughput even under stringent 1.0~ms coherence times. These findings establish contention-aware decentralized routing as a scalable foundation for multi-tenant quantum networks, with direct applicability to emerging quantum data center and distributed quantum computing environments.

\end{abstract}

\section{Introduction}
\label{sec:introduction}

The scalability of multi-tenant quantum networks hinges on their ability to support concurrent entanglement requests from multiple users~\cite{kimble2008quantum}. While our previous protocols---Asynchronous Entanglement Routing (AER)~\cite{yang2024asynchronous} and Multipartite Asynchronous Entanglement Routing (MAER)~\cite{tian2026asynchronous}---have demonstrated efficient asynchronous routing for single-pair and multipartite sessions and outperform existing synchronous methods, they lack mechanisms to resolve local resource contention when multiple source--destination (S--D) pairs compete for shared repeater nodes~\cite{briegel1998quantum} or entanglement links.

In such multi-tenant scenarios, a naive extension of single-request protocols like AER and group-based protocols like MAER---which assume isolated entanglement sessions---can result in resource deadlock. Specifically, when multiple S--D pairs concurrently select paths that share a common repeater node, they may each reserve a qubit for swapping~\cite{pan1998experimental} without coordination. Since quantum memory is finite and non-bufferable, and neither AER nor MAER provides a mechanism for distributed resource arbitration, all competing requests may block indefinitely, leading to mutual failure. This limitation is acknowledged in both works, where uncoordinated resource contention degrades both throughput and end-to-end fidelity.

A fundamental challenge arises: how can a node make intelligent routing decisions using only local knowledge, while simultaneously accounting for (i) probabilistic Bell-State Measurement (BSM) success, (ii) qubit decoherence due to storage and gate noise, and (iii) real-time competition for limited quantum memory? Synchronous protocols sidestep this by assuming global coordination~\cite{shi2024concurrent}; while effective, such coordination incurs scheduling overhead that grows with network size and user concurrency, and requires all nodes to share a common time reference. Conversely, existing distributed asynchronous approaches---such as shortest-path routing over instantaneous link graphs~\cite{pant2019routing}---do not account for real-time resource availability. When multiple requests independently select overlapping paths, the resulting contention induces queuing delays that exhaust the limited coherence time of qubits and degrade overall throughput.

This paper presents the RADAR-Q protocol to address these challenges. Unlike prior works such as AER~\cite{yang2024asynchronous} and MAER~\cite{tian2026asynchronous}, which treat multi-tenant support as an add-on coordination layer, RADAR-Q embeds resource competition directly into the path selection metric. Specifically, it introduces a rank-differential-based objective that jointly optimizes (i) path potential (via hop count $d_{\text{hop}}$ and fidelity $F_v$), (ii) real-time link availability, and (iii) the number of intermediate BSM operations along the path (i.e., BSM depth)---a quantity that governs both the aggregate success probability ($\propto q^k$) and the cumulative decoherence exposure. This metric is evaluated locally at each node using only its Destination-Oriented Directed Acyclic Graph (DODAG)-maintained rank hierarchy and immediate neighbor state, eliminating any need for global topology knowledge or centralized scheduling. By selecting paths with high rank differentials and low contention (quantified by memory occupancy $Q_v$), RADAR-Q proactively resolves conflicts during the routing decision itself.

Our key insight is that contention-awareness, not mere asynchronicity, is the cornerstone of scalable multi-tenant quantum routing. This reframing transforms RADAR-Q from an incremental extension of AER/MAER into a conceptually distinct protocol: whereas prior protocols focus on preserving entanglements across rounds, RADAR-Q optimizes their concurrent utilization under resource competition. For instance, by identifying the Nearest Common Ancestor (NCA) as a localized swapping point, RADAR-Q establishes shorter paths compared to root-centric approaches~\cite{pant2019routing}, reducing both the BSM chain length and the time qubits spend in memory.

Our simulation results demonstrate that RADAR-Q achieves substantially higher entanglement rates compared to both a synchronized NCA-based baseline and an asynchronous root-centric baseline. While global coordination yields the highest fidelity, RADAR-Q sustains a stable end-to-end fidelity well above the 0.5 distillation threshold~\cite{bennett1996purification}. It strategically accepts marginal increases in memory exposure to avoid structural bottlenecks, thereby preserving physical usability for all generated pairs. In stark contrast, naive asynchronous protocols suffer catastrophic fidelity collapse below the distillation threshold and severe fairness degradation (Jain's fairness index~\cite{jain1984quantitative} $\ll 100\%$), precluding their deployment in multi-tenant environments where service predictability and entanglement validity are non-negotiable.

The problem of multi-tenant resource contention arises in any shared quantum infrastructure---from near-term Noisy Intermediate-Scale Quantum (NISQ) testbeds to future error-corrected quantum data centers and distributed quantum computing platforms. In this work, we model the network as a resource-constrained graph where users concurrently compete for finite quantum memory at repeater nodes (i.e., intermediate network nodes equipped with quantum memories and BSM capabilities) and entanglement links, following the abstraction in~\cite{shapourian2025quantumdatacenterinfrastructures}. We evaluate RADAR-Q on two representative topologies: a $10 \times 10$ grid, which provides a regular fabric with uniform connectivity, and a random graph ($N=100$, average degree~4), which captures irregular, heterogeneous link structures. These general topologies allow us to isolate RADAR-Q's core contribution---contention-aware resource arbitration---without presupposing a specific physical architecture, while remaining directly applicable to structured quantum data center interconnects as they mature.

The remainder of this paper is organized as follows. Section~\ref{sec:preliminaries} defines the network model and routing challenges. Section~\ref{sec:RADAR-Q} details the RADAR-Q protocol, including its contention-aware routing metric and DODAG-based path discovery. Section~\ref{sec:simulation} presents simulation settings and results with analysis. Finally, Section~\ref{sec:conclusion} concludes the paper.

\section{Preliminaries}
\label{sec:preliminaries}

This section gives background for the proposed routing scheme. We first explain entanglement generation and swapping. After that, we describe existing synchronous routing protocols and their main problems. Finally, we define the multi-tenant entanglement routing problem.

\subsection{Entanglement Generation and Swapping}
We assume a heralded entanglement generation process in which successful link creation is confirmed by a classical herald signal~\cite{shapourian2025quantumdatacenterinfrastructures}. One concrete realization is the Emitter--Scatter (E-S) scheme: a source node generates a locally entangled pair, retains one stationary qubit $Q_A$, and transmits a flying qubit $Q_F$ entangled with $Q_A$. Upon successful reception, the receiver stores the photon-mediated state as $Q_B$ and returns a classical herald confirming link establishment. Other heralded schemes---such as Emitter--Emitter or Scatter--Scatter protocols---produce equivalent link-level entanglement; the routing layer of RADAR-Q is agnostic to the specific generation mechanism and requires only that each link reports a success probability $p_{uv}$ and an initial fidelity $F_{uv}$. To extend connectivity across multiple hops, intermediate repeater nodes perform Bell-State Measurements (BSMs) that project locally entangled segments onto end-to-end EPR pairs ($Q_A$,$Q_B$). This physical-layer foundation enables RADAR-Q's contention-aware routing under asynchronous operation.

\subsection{Existing Synchronous Approaches}
Most current routing protocols use synchronized time slots, such as~\cite{pant2019routing}. Time is divided into equal-size slots. There are two steps. First, adjacent nodes generate direct-link entanglement pairs in one slot. This creates an ``instant topology.'' Second, the network tries to build end-to-end entanglement by swapping along pre-chosen paths based on the instant topology in the next slot.

However, the protocol enforces a strict boundary between slots: at the end of each one, all entangled pairs---used or unused---are discarded. This is necessary because entangled states decay quickly and cannot be copied or buffered across rounds. Consequently, any entanglement not utilized within its slot is lost, wasting resources and reducing throughput under contention. Furthermore, maintaining strict global synchronization across a large-scale network introduces significant signaling overhead and hardware complexity~\cite{shi2024concurrent,yang2024asynchronous}, making such centralized coordination increasingly costly as the network grows. These limitations motivate our asynchronous, locality-first approach.

\subsection{Problem Statement}
In multi-tenant quantum networks, many users often share the same repeater nodes. This causes resource contention. Entangled states cannot be copied due to the no-cloning theorem~\cite{Wootters1982Single}. They also lose quality over time because of $T_2$ decoherence---the transverse relaxation time that governs how quickly a qubit's phase coherence decays~\cite{osti_1656632}. Therefore, they cannot be stored for long or queued like classical data.

If a request waits too long, its entanglement may expire. This lowers both the success rate and the fidelity of the final link. To avoid this, we need a routing protocol that handles contention well. Its goal is to find paths that reduce contention and maximize the number of successful end-to-end EPR pairs per second.

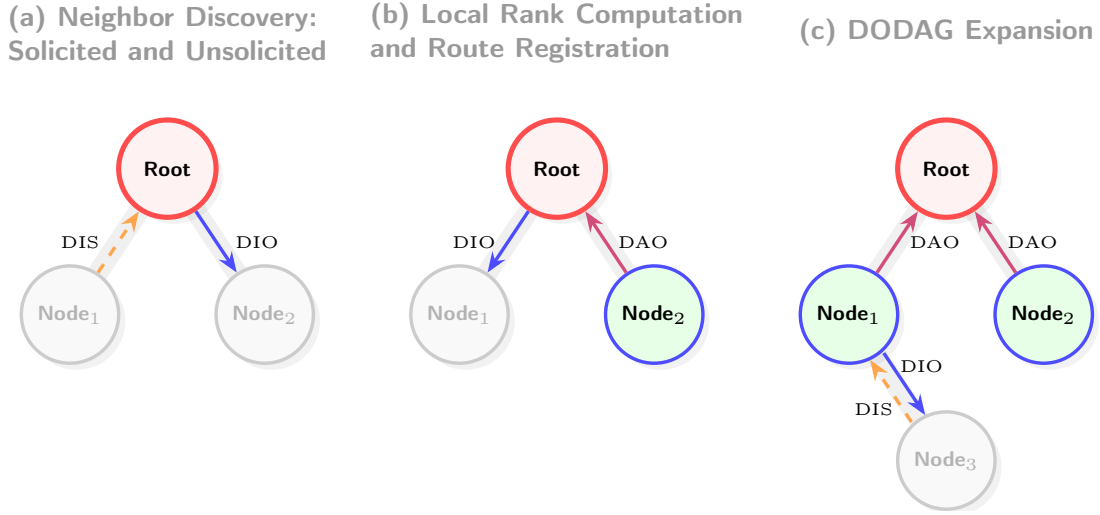
\begin{figure}[htbp]
    \centering
    \begin{tikzpicture}[
    node_base/.style={circle, line width=1pt, minimum size=10mm, font=\sffamily\fontsize{6pt}{7pt}\selectfont\bfseries, drop shadow={opacity=0.1}},
    rootnode/.style={node_base, draw=red!70, fill=red!5, line width=1.5pt},
    joined_node/.style={node_base, draw=blue!70, fill=green!10},
    unjoined_node/.style={node_base, draw=gray!40, fill=gray!5, text=gray!60},
    dis_msg/.style={-{Stealth[scale=0.8]}, draw=orange!70, line width=1pt, dashed}, 
    dio_msg/.style={-{Stealth[scale=0.8]}, draw=blue!70, line width=1pt},        
    dao_msg/.style={-{Stealth[scale=0.8]}, draw=purple!70, line width=1pt}, 
    bundle/.style={draw=gray!12, line width=6pt, line cap=round},
    panel_title/.style={
        font=\sffamily\bfseries\fontsize{8pt}{10pt}\selectfont, 
        color=gray!80
    }
]

    \begin{scope}[local bounding box=phase_a]
        \node[panel_title, align=left] at (1,1.4) {(a) Neighbor Discovery:\\Solicited and Unsolicited};
        
        \node[rootnode] (R) at (1,0) {Root};
        \node[unjoined_node] (N1) at (0,-1.5) {Node$_1$};
        \node[unjoined_node] (N2) at (2,-1.5) {Node$_2$};

        \begin{pgfonlayer}{background}
            \draw[bundle] (R.center) -- (N1.center);
            \draw[bundle] (R.center) -- (N2.center);
        \end{pgfonlayer}
        
        \draw[dis_msg] (N1) -- (R) node[midway, left, font=\tiny, xshift=-2pt] {DIS};
        
        \draw[dio_msg] (R) -- (N2) node[midway, right, font=\tiny, xshift=2pt] {DIO};
    \end{scope}

    \begin{scope}[xshift=4cm]
        \node[panel_title, align=left] at (1,1.4) {(b) Local Rank Computation \\ and Route Registration};
        
        \node[rootnode] (R) at (1,0) {Root};
        \node[unjoined_node] (N1) at (0,-1.5) {Node$_1$};
        \node[joined_node] (N2) at (2,-1.5) {Node$_2$};

        \begin{pgfonlayer}{background}
            \draw[bundle] (R.center) -- (N1.center);
            \draw[bundle] (R.center) -- (N2.center);
        \end{pgfonlayer}

        \draw[dio_msg] (R) -- (N1) node[midway, left, font=\tiny] {DIO};
        
        \draw[dao_msg] (N2) -- (R) node[midway, right, font=\tiny] {DAO};
    \end{scope}

    \begin{scope}[xshift=8cm]
        \node[panel_title, align=left] at (1,1.4) {(c) DODAG Expansion};
        
        \node[rootnode] (R) at (1,0) {Root};
        \node[joined_node] (N1) at (0,-1.5) {Node$_1$}; 
        \node[joined_node] (N2) at (2,-1.5) {Node$_2$};
        \node[unjoined_node] (N3) at (1,-3) {Node$_3$};
        
        \begin{pgfonlayer}{background}
            \draw[bundle] (R.center) -- (N1.center);
            \draw[bundle] (R.center) -- (N2.center);
            \draw[bundle] (N1.center) -- (N3.center);
        \end{pgfonlayer}

        \draw[dao_msg] (N1) -- (R) node[midway, right, font=\tiny] {DAO};
        \draw[dao_msg] (N2) -- (R) node[midway, right, font=\tiny] {DAO};

        \draw[dio_msg, transform canvas={xshift=2pt, yshift=1pt}] (N1) -- (N3) 
            node[midway, above, xshift=5pt, font=\tiny] {DIO};
        
        \draw[dis_msg, transform canvas={xshift=-2pt, yshift=-1pt}] (N3) -- (N1) 
            node[midway, below, xshift=-5pt, font=\tiny] {DIS};
    \end{scope}

\end{tikzpicture}
    \caption{Distributed signaling for DODAG maintenance. Neighbor discovery and rank propagation via DIS, DIO, and DAO messages in an asynchronous environment.}
    \label{fig.dodag_expansion}
\end{figure}

\section{RADAR-Q: Contention-Aware Asynchronous Entanglement Routing}
\label{sec:RADAR-Q}

This section describes the RADAR-Q protocol. We first define our network model. Then we explain how nodes form and maintain a logical DODAG for topology discovery. Next, we show the advantages of path selection using the NCA based on the DODAG structure. Finally, we present the routing algorithm and the contention-aware metric used to select optimal paths.

\subsection{Network Topology Model}
We consider a physical quantum network topology modeled as an undirected graph $\mathcal{G} = (\mathcal{V}, \mathcal{E})$, where $\mathcal{V}$ is the set of nodes (users and repeaters) and $\mathcal{E}$ represents bidirectional classical-quantum hybrid channels. Each channel supports both qubit transmission and classical communication. Each node $v \in \mathcal{V}$ is equipped with a finite quantum memory of size $M_v$ qubits and a local BSM device capable of performing entanglement swapping between any two qubits stored in its memory. A direct link $(u, v) \in \mathcal{E}$ can generate entangled pairs at a success probability $p_{uv}$ per attempt, with a fidelity $F_{uv}$ that decays due to two primary noise sources in the NISQ era: (i) storage noise, which accumulates exponentially during the coherence time $T_{\text{CO}}$, and (ii) gate noise introduced by each BSM operation, which itself succeeds with a probability $q$. To ensure physical feasibility, the routing state at any node $v$ is constrained by its memory size $M_v$. Let \( Q_v \in \{0,1,\dots,M_v\} \) denote the current quantum memory occupancy at node \( v \), i.e., the number of qubits currently storing entangled states. The memory utilization ratio is then \( Q_v / M_v \in [0,1] \). Under a best-effort delivery model, we assign uniform weights to all requests $\langle s, d \rangle \in R$.

\begin{figure}[htbp]
    \centering
    \begin{tikzpicture}[
    gridnode/.style={circle, draw=blue!70, fill=blue!5, line width=1.2pt, minimum size=10mm, font=\sffamily\fontsize{8pt}{9pt}\selectfont\bfseries, drop shadow={opacity=0.1}},
    root_g/.style={circle, draw=red!70, fill=red!5, line width=1.6pt, minimum size=10mm, font=\sffamily\fontsize{8pt}{9pt}\selectfont\bfseries, drop shadow={opacity=0.2}},
    user1_g/.style={circle, fill=cyan!15, draw=cyan!70, line width=1.2pt, minimum size=10mm, font=\sffamily\fontsize{13pt}{9pt}\selectfont\bfseries, drop shadow={opacity=0.1}},
    user2_g/.style={circle, fill=green!15, draw=green!70, line width=1.2pt, minimum size=10mm, font=\sffamily\fontsize{13pt}{9pt}\selectfont\bfseries, drop shadow={opacity=0.1}},
    tree_node/.style={circle, draw=blue!70, fill=blue!5, line width=1pt, minimum size=10mm, font=\sffamily\fontsize{8pt}{7pt}\selectfont\bfseries, drop shadow={opacity=0.1}},
    root_t/.style={circle, draw=red!70, fill=red!5, line width=1.5pt, minimum size=10mm, font=\sffamily\fontsize{8pt}{7pt}\selectfont\bfseries},
    user_t/.style={circle, fill=cyan!10, draw=cyan!70, line width=1pt, minimum size=10mm, font=\sffamily\fontsize{13pt}{7pt}\selectfont\bfseries},
    bundle_sleeve/.style={draw=gray!12, line width=8pt, line cap=round},
    nca_path_1/.style={draw=purple!70, line width=2.5pt, double=purple!5, double distance=1.5pt, opacity=0.9},
    nca_path_2/.style={draw=green!60!black, line width=2.5pt, double=green!5, double distance=1.2pt, opacity=0.9},
    root_detour/.style={draw=red!50, line width=1.2pt, -{Stealth[scale=1.0, fill=red!50]}},
    fig_title/.style={font=\sffamily\bfseries\LARGE\color{gray!80}},
    mapping_arrow/.style={
        -{Stealth[scale=1.5, fill=gray!40]}, 
        line width=1.5pt, 
        gray!40, 
        double, 
        double distance=5pt
    }
]

    \begin{scope}[local bounding box=grid_top]
        \node[fig_title] at (3.6, 1.5) {Physical Grid Topology};
        
        \begin{pgfonlayer}{background}
            \foreach \ix in {0,1,2,3,4} {
                \foreach \iy in {0,1,2,3,4} {
                    \ifnum\ix<4 \draw[bundle_sleeve] (\ix*1.8, -\iy*1.8) -- (\ix*1.8+1.8, -\iy*1.8); \fi
                    \ifnum\iy<4 \draw[bundle_sleeve] (\ix*1.8, -\iy*1.8) -- (\ix*1.8, -\iy*1.8-1.8); \fi
                }
            }
        \end{pgfonlayer}

        \node[root_g] (GR) at (3.6,-3.6) {Root};
        \node[gridnode] (GR1) at (1.8,-3.6) {NCA$_1$};
        \node[gridnode] (GR2) at (1.8,-1.8) {R2}; 
        \node[gridnode] (GR3) at (1.8,-5.4) {R3}; 
        \node[gridnode] (GNCA1) at (5.4,-3.6) {R1};
        \node[gridnode] (GNCA2) at (5.4,-1.8) {NCA$_2$};
        
        \node[user1_g] (GS1) at (0,-1.8) {s$_1$};
        \node[user1_g] (GD1) at (0,-5.4) {d$_1$};
        \node[user2_g] (GS2) at (5.4, 0) {s$_2$};
        \node[user2_g] (GD2) at (7.2, -1.8) {d$_2$};

        \begin{pgfonlayer}{L1}
            \draw[root_detour] (GS1) -- (GR2) -- (GR1) -- (GR);
            \draw[root_detour] (GD1) -- (GR3) -- (GR1) -- (GR);

            \draw[root_detour] (GS2) -- (GNCA2) -- (GNCA1) -- (GR);
            \draw[root_detour] (GD2) -- (GNCA2) -- (GNCA1) -- (GR);

            \draw[nca_path_1] (GS1.center) -- (GR2.center) -- (GR1.center) -- (GR3.center) -- (GD1.center);
            \draw[nca_path_2] (GS2.center) -- (GNCA2.center) -- (GD2.center);
        \end{pgfonlayer}

    \end{scope}


   
    \begin{scope}[xshift=15.5cm] 
        \node[fig_title] at (0, 1.5) {Logical DODAG Topology};

        \begin{pgfonlayer}{background}
            \foreach \r/\y in {0/-1, 1/-3, 2/-5, 3/-6.9} {
                
                \draw[gray!50, dashed, line width=1pt] (-4.3, \y) -- (4.5, \y);
               
                \node[anchor=west, font=\sffamily\small\bfseries, color=gray!50] at (4.6, \y) {Rank-\r};
            }
        \end{pgfonlayer}

        \node[root_t] (R) at (0,0) {Root}; 
        
        \node[tree_node] (N1L) at (-2,-2) {NCA$_{1}$}; 
        \node[tree_node] (N1R) at (2,-2) {R1};
        
        \node[tree_node] (R2) at (-3,-4) {R2}; 
        \node[tree_node] (R3) at (-1,-4) {R3};
        \node[tree_node] (NCA2) at (1.5,-4) {NCA$_2$};
        
        \node[user_t] (S1) at (-3.5,-6) {s$_1$}; 
        \node[user_t] (D1) at (-0.5,-6) {d$_1$};
        \node[user_t, fill=green!10, draw=green!70] (S2) at (1,-6) {s$_2$};
        \node[user_t, fill=green!10, draw=green!70] (D2) at (3.5,-6) {d$_2$};

        \begin{pgfonlayer}{background}
            \foreach \u/\v in {R/N1L, R/N1R, N1L/R2, N1L/R3, N1R/NCA2, R2/S1, R3/D1, NCA2/S2, NCA2/D2}
                \draw[bundle_sleeve] (\u.center) -- (\v.center);
        \end{pgfonlayer}

        \begin{pgfonlayer}{L1}
           
            \draw[root_detour] (S1) -- (R2) -- (N1L) -- (R);
            \draw[root_detour] (D1) -- (R3) -- (N1L) -- (R);
            \draw[root_detour] (S2) -- (NCA2) -- (N1R) -- (R);
            \draw[root_detour] (D2) -- (NCA2) -- (N1R) -- (R);

            \draw[nca_path_1] (S1.center) -- (R2.center) -- (N1L.center) -- (R3.center) -- (D1.center);
            \draw[nca_path_2] (S2.center) -- (NCA2.center) -- (D2.center);
        \end{pgfonlayer}

        \node[draw=gray!40, dashed, inner sep=8pt, rounded corners=5pt, fit=(N1L) (R2) (R3) (S1) (D1)] (Bound) {};
        \node[below=2pt of Bound, font=\sffamily\small\color{gray}] {$\bullet$ Bounded Local Knowledge};
    \end{scope}

    \draw[mapping_arrow] (7.5, -3) -- (11, -3)
        node[midway, above, font=\sffamily\bfseries\small, color=gray!80, yshift=10pt] {DODAG MAPPING};
        
    
    \begin{scope}[yshift=-9.5cm, xshift=2.5cm] 
        \draw[draw=gray!40, fill=gray!2, rounded corners=5pt, line width=0.8pt] 
            (-1.8, 1.3) rectangle (18, -1.2);

        \node[font=\sffamily\bfseries\large, color=gray!90, anchor=west] at (-1.5, 0.8) {Legend:};

        \node[circle, fill=cyan!15, draw=cyan!70, line width=1pt, minimum size=7mm, inner sep=0pt] (leg1) at (0.8, 0.3) {};
        \node[font=\sffamily\normalsize, anchor=west] at (1.3, 0.3) {User Pair 1 ($s_1, d_1$)};
        
        \node[circle, fill=green!15, draw=green!70, line width=1pt, minimum size=7mm, inner sep=0pt] (leg2) at (5.8, 0.3) {};
        \node[font=\sffamily\normalsize, anchor=west] at (6.3, 0.3) {User Pair 2 ($s_2, d_2$)};
        
        \node[circle, fill=blue!5, draw=blue!70, line width=1pt, minimum size=7mm, inner sep=0pt] (leg3) at (10.8, 0.3) {};
        \node[font=\sffamily\normalsize, anchor=west] at (11.3, 0.3) {Repeater / NCA};

        \draw[draw=purple!70, line width=2.8pt, double=purple!5, double distance=1.6pt] (0.4, -0.6) -- (1.4, -0.6);
        \node[font=\sffamily\normalsize, anchor=west] at (1.6, -0.6) {Pair 1 Optimized Path};
        
        \draw[draw=green!60!black, line width=2.8pt, double=green!5, double distance=1.3pt] (6.5, -0.6) -- (7.5, -0.6);
        \node[font=\sffamily\normalsize, anchor=west] at (7.7, -0.6) {Pair 2 Optimized Path};
        
        \draw[draw=red!50, line width=1.4pt, -{Stealth[scale=1.2, fill=red!50]}] (12.3, -0.6) -- (13.3, -0.6);
        \node[font=\sffamily\normalsize, anchor=west] at (13.5, -0.6) {Entanglement Links};
        
    \end{scope}

\end{tikzpicture}
    \caption{Structural mapping from a physical grid topology (left) to the corresponding logical DODAG (right). Node colors indicate roles: cyan for user pair~1 ($s_1$, $d_1$), green for user pair~2 ($s_2$, $d_2$), blue for repeater nodes, and red for the DODAG root. Red dashed arrows show root-centric default paths; colored solid lines show RADAR-Q's NCA-optimized paths. The grid serves as an illustrative example; RADAR-Q operates on arbitrary graph topologies.}
    \label{fig:dodag_mapping}
\end{figure}
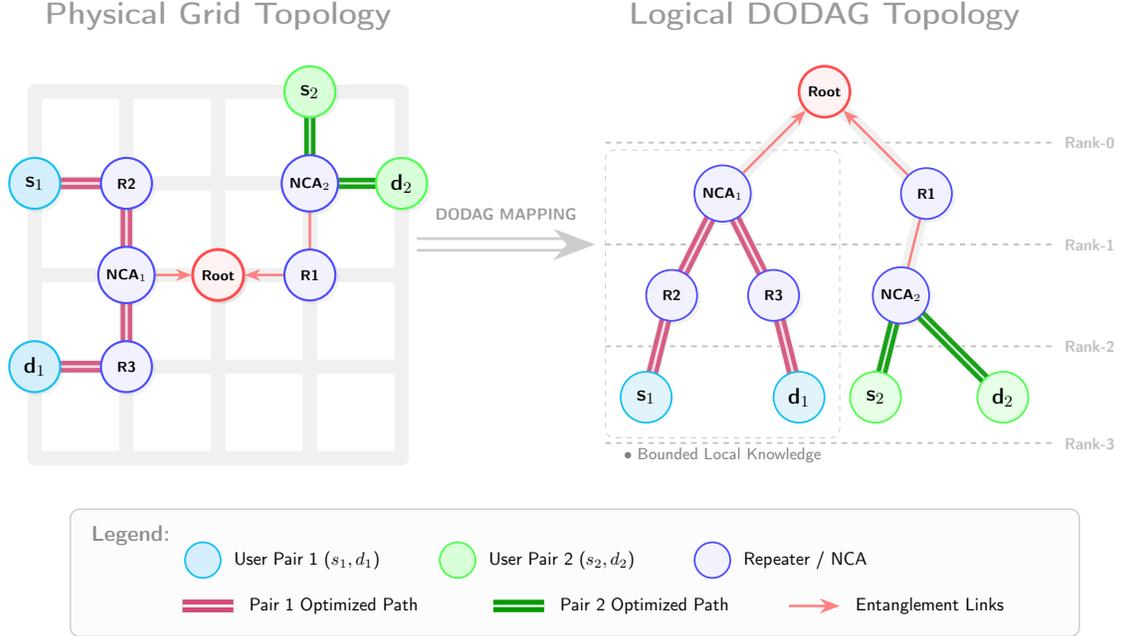

\subsection{DODAG Formation and Maintenance}
RADAR-Q adopts the DODAG structure from the Routing Protocol for Low-Power and Lossy Networks (RPL) protocol~\cite{rfc6550}. A single root node anchors the DODAG, serving as the reference point for rank computation. All other nodes self-organize by calculating their rank using only local information and advertisements received from neighbors. The DODAG is constructed and maintained via three classical control message types, extended in RADAR-Q with quantum-specific parameters:
\begin{itemize}
    \item DODAG Information Object (DIO): multicast downward to advertise a node's rank, DODAG version, and quantum metrics including average link fidelity $F_v$ and memory utilization $Q_v$.
    \item DODAG Information Solicitation (DIS): multicast by a node to solicit DIO messages from neighbors for joining or state refresh.
    \item Destination Advertisement Object (DAO): sent unicast upward to the selected parent after parent selection. DAOs propagate reachability information toward the root. It allows storing-mode ancestors to cache downward routes to descendants.
\end{itemize}
Figure~\ref{fig.dodag_expansion} illustrates this distributed signaling sequence.

\subsection{Rank Definition and Loop-Free Guarantee}
The rank of node $v$ in a DODAG is defined as a composite metric that prioritizes path length while penalizing poor quantum quality or resource contention:

\begin{equation}
  \text{rank}(v) = d_{\text{hop}}(v) + \underbrace{\frac{\alpha (1 - F_v) + \beta (Q_v / M_v)}{\alpha + \beta}}_{\in [0, 1)}
  \label{eq:rank_def}
\end{equation}

where
\begin{itemize}
    \item $d_{\text{hop}}(v) \in \mathbb{N}_0$ is the hop distance from $v$ to the root,
    \item $F_v \in [0,1]$ is the average fidelity over $v$'s links to candidate parents,
    \item $Q_v / M_v \in [0,1]$ is the fraction of occupied quantum memory slots at $v$,
    \item $\alpha, \beta > 0$ are tunable weights balancing fidelity loss versus memory load.
\end{itemize}

The fractional term is strictly bounded in $[0, 1)$, ensuring that a node's rank is always less than its parent's rank plus one:
\[
\text{rank}(v) < \text{rank}(\text{parent}(v)) + 1
\]
This strict monotonic increase along any downward path prevents routing loops, while still allowing quantum-aware tie-breaking within the same hop distance. Each node selects as its preferred parent the neighbor advertising the lowest rank. Combined with DAO-based reachability caching at ancestors, this mechanism enables every node to locally trace a consistent upward path toward potential NCAs for any S--D pair and bypass the global knowledge need. Figure~\ref{fig:dodag_mapping} exemplifies the mapping from a physical grid topology to the resulting logical DODAG.

Note that Eq.~\ref{eq:rank_def} governs the DODAG construction phase (line~1 of Algorithm~\ref{alg:RADAR-Q}): when \textsc{UpdateDODAG} is invoked, each node recomputes its rank using Eq.~\ref{eq:rank_def} based on the latest fidelity and memory advertisements from neighbors. The resulting rank hierarchy determines parent selection and, consequently, the set of candidate upward paths available for routing. The contention-aware path selection metric (Eq.~\ref{eq:contention-aware_metric}) then operates over these DODAG-derived paths.

\subsection{NCA-Centric Path Discovery and Selection}
In many existing protocols, entanglement paths are computed as shortest paths over instantaneous link graphs~\cite{pant2019routing}, without considering how multiple concurrent requests interact at shared nodes. When these shortest paths converge at common repeaters, contention emerges. RADAR-Q addresses this by adopting an NCA-centric approach. As illustrated in the left portion of Figure~\ref{fig:dodag_mapping}, RADAR-Q identifies a localized swapping point closer to the users. For example, the request pair $\langle s_2, d_2 \rangle$ can complete entanglement swapping at $NCA_2$ instead of traversing the entire path to the root. For any S--D pair $\langle s, d \rangle$, entanglement swapping is localized at their NCA node, defined as the deepest common ancestor in the DODAG tree. The resulting path consists of two upward segments: $s \to \text{nca}(s,d)$ followed by $d \to \text{nca}(s,d)$. Let $k' = d_{\text{hop}}(s, \text{nca}) + d_{\text{hop}}(d, \text{nca})$ denote the path length in hops. By construction, $k' \leq k$, where $k$ is the length of the root-routed path $s \to \text{root} \to d$. Equality holds only when the root is the sole common ancestor.

\begin{algorithm}[t]
\caption{RADAR-Q: Contention-Aware Routing}
\label{alg:RADAR-Q}
\begin{algorithmic}[1]
\Require Local DODAG view $G_{\text{inst}}$, concurrent requests $\mathcal{R}=\{\langle s_i,d_i\rangle\}$
\Ensure Routing status and established paths
\State $G_{\text{inst}} \gets \text{UpdateDODAG}(G_{\text{inst}})$ \Comment{Ranks via Eq.~\ref{eq:rank_def}}
\State Sort $\mathcal{R}$ by $d_{\text{hop}}(\text{nca}(s,d))$ descending \Comment{Locality-first}
\For{each $\langle s,d\rangle \in \mathcal{R}$}
    \State $\mathcal{P} \gets \text{GenerateCandidatePaths}(s,d,G_{\text{inst}})$
    \If{$\mathcal{P} = \emptyset$} \textbf{continue} \EndIf
    \State $p^* \gets \arg\max_{p\in\mathcal{P}} \frac{d_{\text{hop}}(\text{nca}(s,d))+1}{1+\max_{e\in p}(1-\text{avail}(e))}$ \Comment{Eq.~\ref{eq:contention-aware_metric}}
    \If{$\text{AttemptSwaps}(p^*)$ succeeds}
        \State $\text{LogSuccess}(\langle s,d\rangle, p^*)$
    \Else
        \State $\text{NotifyFailure}(p^*)$; $\text{TriggerLocalizedUpdate}(p^*)$
    \EndIf
\EndFor
\end{algorithmic}
\end{algorithm}

\subsection{NCA-Centric Routing and Contention Awareness}

Algorithm~\ref{alg:RADAR-Q} implements a contention-aware routing logic operating entirely on local state. Upon receiving concurrent requests $\mathcal{R} = \{\langle s_i, d_i \rangle\}$, RADAR-Q employs a Locality-First Priority mechanism: requests are processed in descending order of NCA depth $d_{\text{hop}}(\text{nca}(s,d))$, prioritizing localized routes that require fewer BSM operations and maximize success probability under finite coherence times.

For each request $\langle s,d \rangle$, the node identifies all common ancestors of $s$ and $d$ within its local DODAG view $G_{\text{inst}}$, sorted by depth (deepest first). Candidate paths are constructed by concatenating upward routes $s \rightarrow \text{nca}$ and $d \rightarrow \text{nca}$; any path containing a saturated link ($\text{avail}(e) = 0$) is discarded. The optimal path $p^*$ maximizes the contention-aware metric:
\begin{equation}
p^* = \arg\max_{p \in \mathcal{P}} \frac{d_{\text{hop}}(\text{nca}(s,d)) + 1}{1 + \max_{e \in p} (1 - \text{avail}(e))},
\label{eq:contention-aware_metric}
\end{equation}
where the numerator rewards locality (deeper NCA $\Rightarrow$ shorter BSM depth) and the denominator penalizes congestion.

Upon BSM failure or decoherence, a lightweight notification triggers localized DODAG updates among affected neighbors, ensuring rapid convergence without global overhead. The $\textsc{FindAllCommonAncestors}$ function operates fully distributed: nodes independently broadcast ancestor lists upward via parent pointers; their first intersection yields the NCA. This reframes multi-tenant routing from reactive conflict management to proactive, metric-driven path optimization.

\section{Simulation and Evaluation}
\label{sec:simulation}

We evaluate RADAR-Q against two architectural extremes---Syn-NCA~\cite{shi2024concurrent} and Asyn-Root~\cite{pant2019routing}---to validate its performance in multi-user quantum networks. These baselines represent the upper bounds of global synchronization and the current standard for distributed root-centric routing, respectively. Our experiments address three primary objectives: (1) Measuring throughput scalability under increasing concurrent demand; (2) Analyzing the trade-off between throughput and end-to-end fidelity; and (3) Assessing the protocol's fairness and robustness in resource-constrained environments.

\subsection{Experimental Setup}
\noindent\textbf{Baselines.} To evaluate the impact of contention-awareness and asynchronicity, we implement the following protocols:
\begin{itemize}
    \item \textbf{Synch-NCA}: A synchronized protocol that utilizes global knowledge for path assignment via Nearest Common Ancestors. It represents an upper bound for fidelity due to its idealized synchronization.
    \item \textbf{Asynch-Root}: A traditional asynchronous protocol where all requests default to the DODAG root, representing the standard distributed approach without contention awareness.
\end{itemize}

\textbf{Network Topologies.} We employ two distinct topologies to model resource contention dynamics in shared quantum networks~\cite{shapourian2025quantumdatacenterinfrastructures}: (i) a $10\times10$ grid network, where repeaters form a regular switching fabric with uniform resource distribution; and (ii) a random graph ($N=100$ nodes, average degree 4) capturing heterogeneous network structures with non-uniform link qualities. These configurations are intentionally chosen to isolate the protocol's core contribution to resource arbitration. Specifically, they allow us to evaluate RADAR-Q's ability to manage concurrent multi-tenant requests under decoherence constraints without presupposing a specific physical architecture, while remaining representative of both structured interconnects (e.g., quantum data centers) and irregular deployments (e.g., metropolitan quantum networks).

\noindent\textbf{Parameters.} We use realistic near-term parameters: link generation probability $p = 0.8$ and BSM success probability $q = 0.9$, consistent with matter-based quantum platforms such as nitrogen-vacancy (NV) centers in diamond~\cite{bradley2019ten, kalb2017entanglement} and trapped-ion systems that support deterministic BSMs. We note that linear-optical photonic BSMs are fundamentally limited to $q \leq 0.5$~\cite{humphreys2018deterministic}; our parameter choice reflects the higher-fidelity regime achievable with matter-qubit repeaters, while the protocol itself is agnostic to the specific $q$ value. The initial fidelity $F_0 = 0.95$, aligning with reported two-qubit entanglement fidelities in near-term hardware~\cite{bradley2019ten}. In scalability tests, we assume $T_{\text{co}} = \infty$ to isolate protocol logic, while robustness tests vary $T_{\text{co}}$ from $1.0$~ms to $\infty$.

The end-to-end fidelity of generated entanglement pairs degrades due to two primary mechanisms: (i) memory decoherence during storage, we model the fidelity decays exponentially with waiting time as $F(t) = \frac{1}{4} + (F_0 - \frac{1}{4}) e^{-t / \tau}$ (with $\tau$ proportional to coherence time $T_{\text{co}}$)~\cite{dur1999quantum}; and (ii) infidelity introduced by each entanglement swapping operation via imperfect BSMs, which multiplicatively reduces fidelity roughly as $\propto q^k$ per attempt (where $k = \text{the number of hops along the path} -1$, and $q$ is the per-swap success probability).

\subsection{Throughput Scalability Analysis}

\begin{figure*}[t]
    \centering
    \begin{subfigure}[b]{0.49\textwidth}
        \centering
        \includegraphics[width=\textwidth]{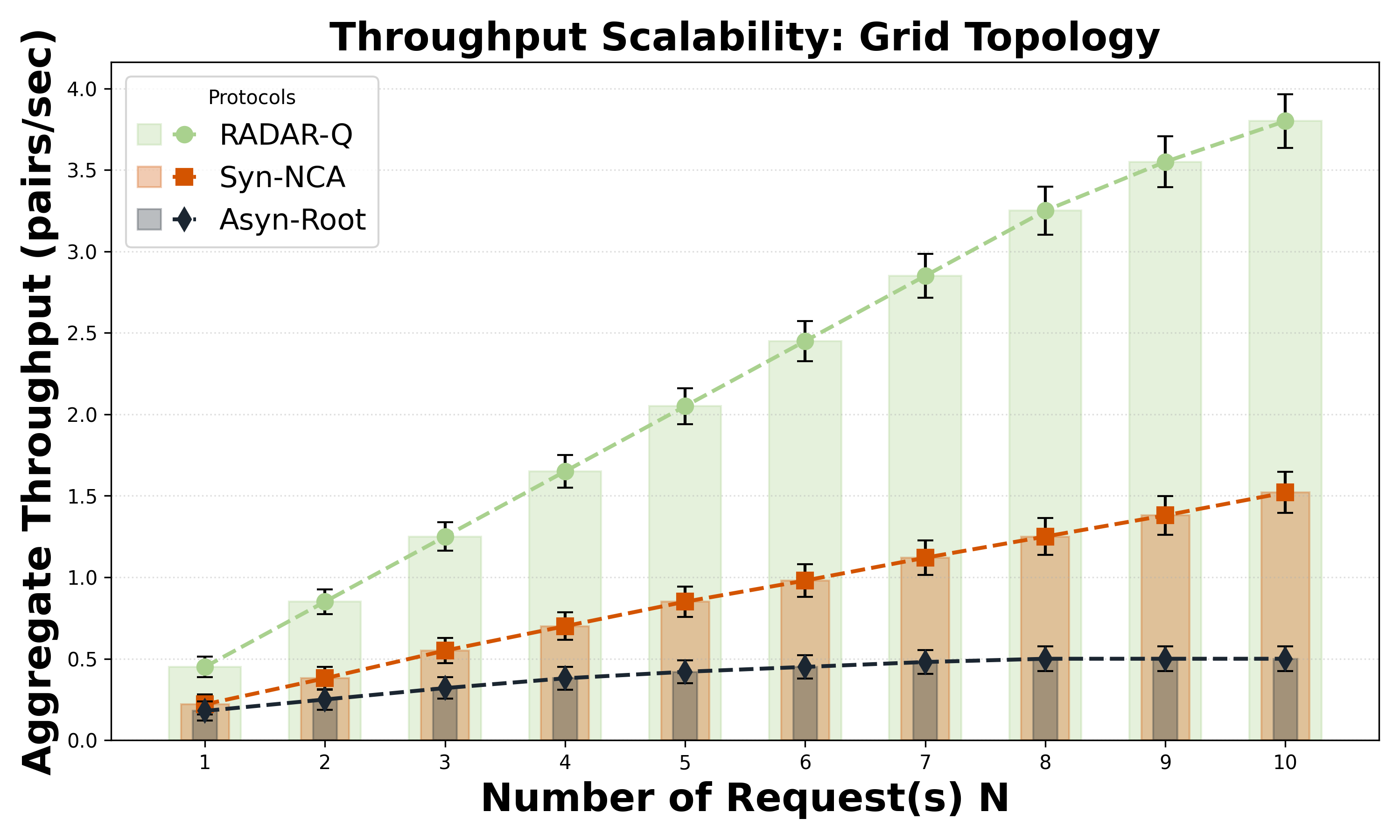}
        \caption{Grid Topology}
        \label{fig:grid_tp}
    \end{subfigure}
    \hfill
    \begin{subfigure}[b]{0.49\textwidth}
        \centering
        \includegraphics[width=\textwidth]{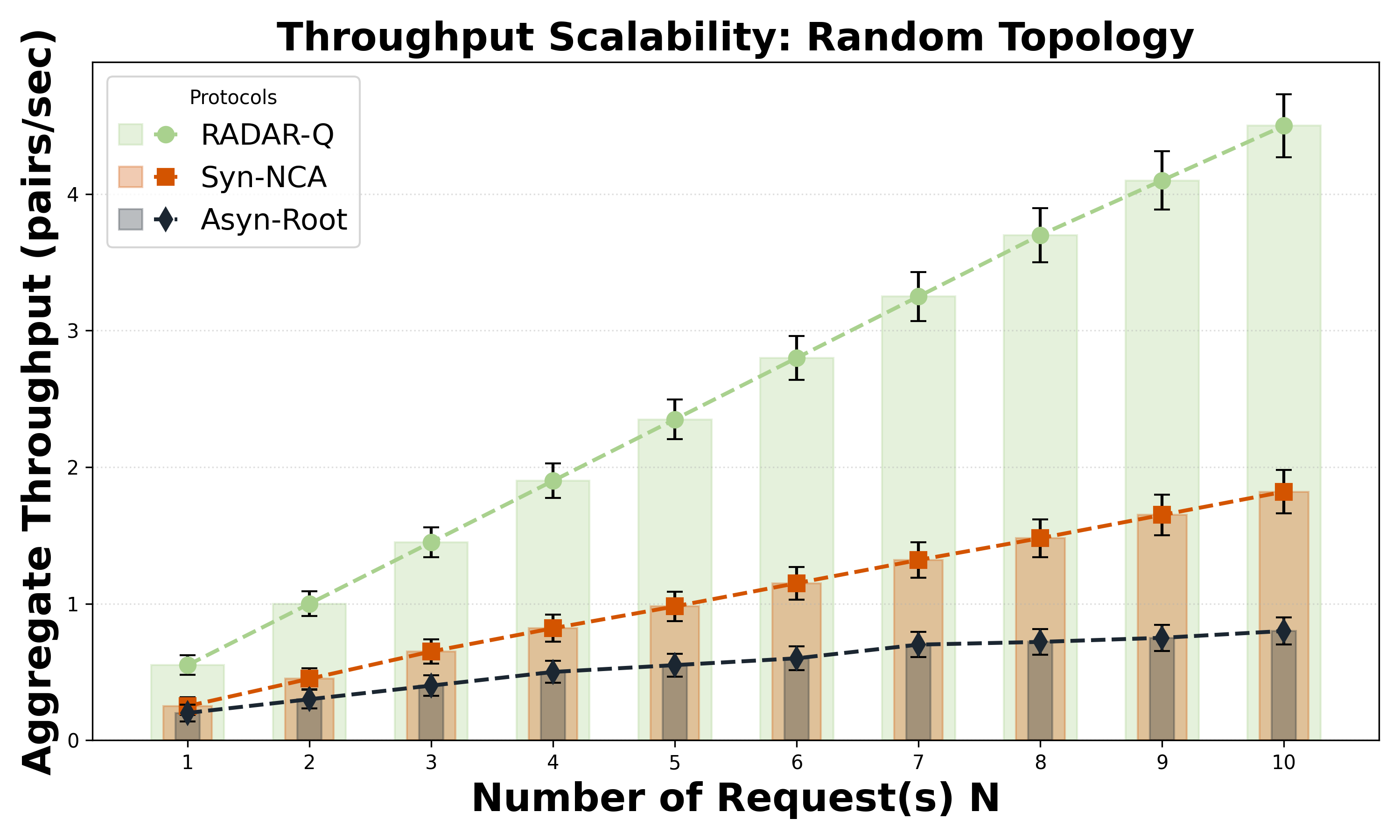}
        \caption{Random Topology}
        \label{fig:rand_tp}
    \end{subfigure}
    \caption{Aggregate throughput scalability ($T_{co} = \infty$). RADAR-Q achieves near-linear growth, outperforming the root-centric baseline by up to 7.6$\times$ in Grid networks.}
    \label{fig:throughput_results}
\end{figure*}

Figures~\ref{fig:grid_tp} and~\ref{fig:rand_tp} demonstrate RADAR-Q’s superior throughput scalability under idealized conditions ($T_{co} = \infty$), effectively isolating the impact of its contention-resolution mechanism. In the grid topology, at $N=10$ concurrent requests, RADAR-Q achieves an aggregate throughput of approximately 3.8 pairs/sec, representing a 2.5$\times$ improvement over Synch-NCA and a 7.6$\times$ gain relative to Asynch-Root. In the random topology, RADAR-Q delivers approximately 4.5 pairs/sec at $N=10$, confirming its robust performance across diverse and less structured network environments.

The near-linear growth of RADAR-Q’s throughput curve validates its ability to distribute load efficiently across the network fabric. By prioritizing NCA-based paths that circumvent the root node, RADAR-Q enables parallel entanglement establishment for multiple user pairs, thereby avoiding the severe resource starvation observed in Asyn-Root. This advantage extends even to Syn-NCA. Despite its global coordination, Syn-NCA incurs significant scheduling overhead and lacks a proactive mechanism to handle path overlapping.

This performance is a direct consequence of RADAR-Q’s contention-aware routing metric (Eq.~\ref{eq:contention-aware_metric}). It proactively identifies low-contention paths using only local state. The divergence between RADAR-Q and the baselines widens significantly as concurrency increases stretching from a modest 2$\times$ gap at $N=1$ to over 7$\times$ at $N=10$. This trend demonstrates that RADAR-Q’s scalability is not merely incremental but fundamentally transformative for supporting high-density multi-tenant quantum communications.

\subsection{Fidelity vs. Throughput Trade-off}

\begin{figure}[t]
    \centering
    \begin{subfigure}[b]{0.49\textwidth}
        \centering
        \includegraphics[width=\textwidth]{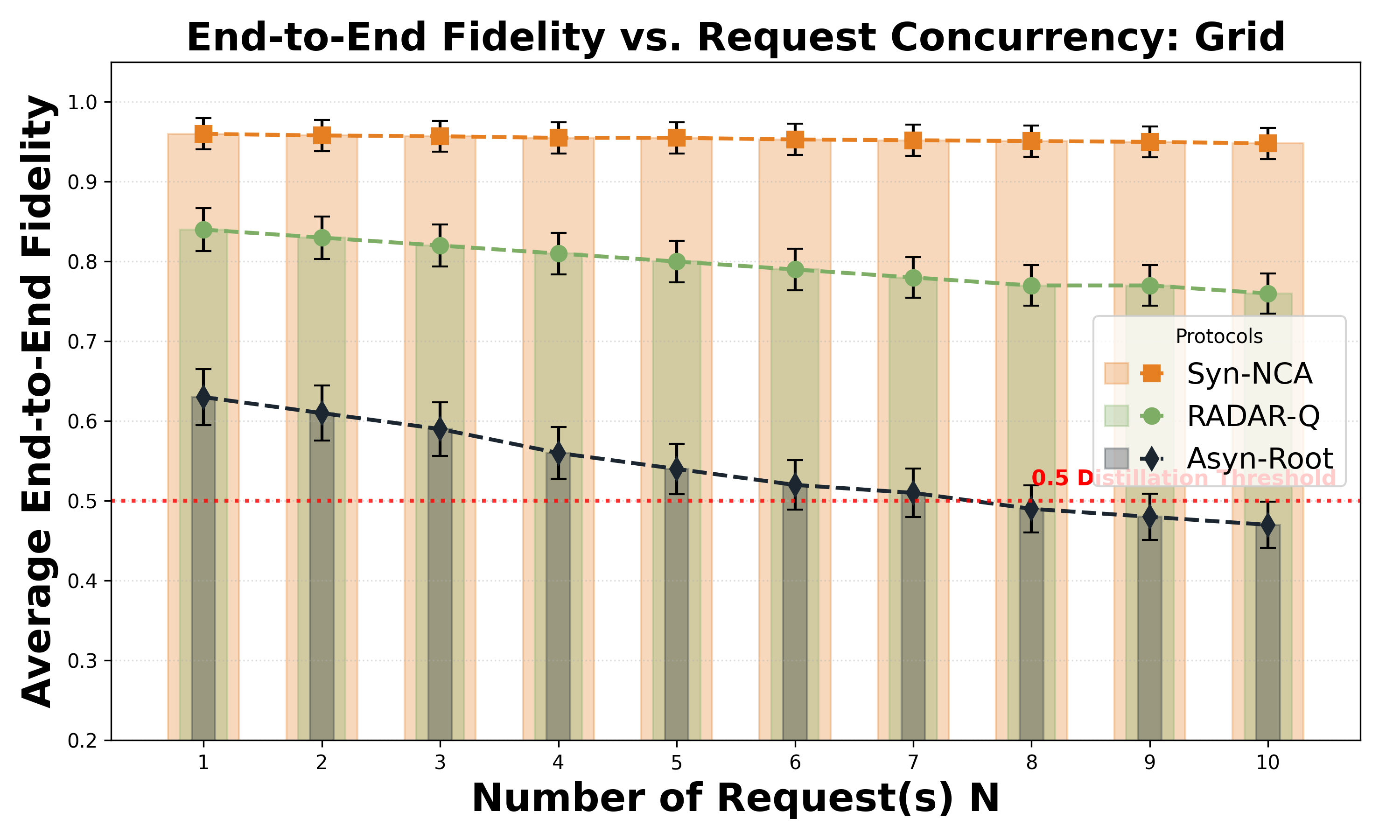}
        \caption{Grid Topology}
        \label{fig:grid_fid}
    \end{subfigure}
    \hfill
    \begin{subfigure}[b]{0.49\textwidth}
        \centering
        \includegraphics[width=\textwidth]{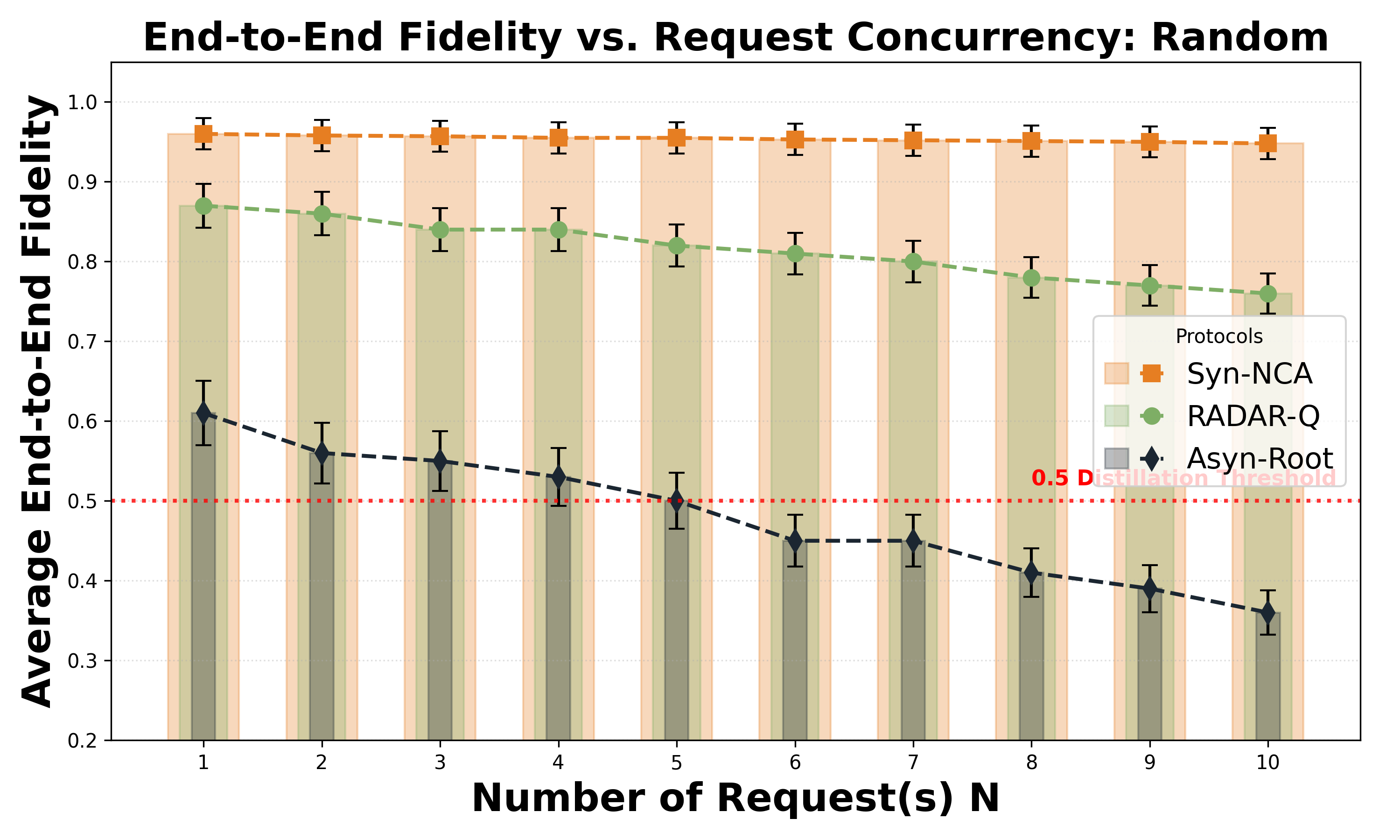}
        \caption{Random Topology}
        \label{fig:rand_fid}
    \end{subfigure}
    \caption{End-to-end fidelity vs. request concurrency. While baseline fidelity collapses below the 0.5 threshold, RADAR-Q maintains physical usability (fidelity $\approx 0.76$) through precision trading.}
    \label{fig:fidelity_results}
\end{figure}

Figures~\ref{fig:grid_fid} and~\ref{fig:rand_fid} present the average end-to-end fidelity as a function of concurrent request count $N$. As anticipated, Syn-NCA maintains the highest fidelity ($\approx 0.95$), benefiting from its globally coordinated path assignment that minimizes link reuse and avoids aged entanglements. In contrast, RADAR-Q exhibits a stable yet slightly lower fidelity profile, hovering around 0.76--0.77 at $N=10$ in both topologies. We view this small fidelity drop as a necessary cost for the performance gains described in Section~\ref{sec:simulation}. By removing the need for global synchronization, RADAR-Q trades a slight increase in memory exposure for the ability to handle many more concurrent users.

While Syn-NCA represents an idealized upper bound, RADAR-Q significantly outperforms the more practical Asyn-Root baseline, drops to 0.48 in Grid and 0.36 in Random. Notably, Asyn-Root fails the 0.5 distillation threshold~\cite{bennett1996purification} under high load, while RADAR-Q maintains graceful degradation well above this limit. This ensures that every pair we generate is actually useful for teleportation or error correction.

The high performance of RADAR-Q comes from keeping the BSM count low. Since the aggregate success probability for a single entanglement attempt scales $\propto q^{k}$, where $q$ is the hardware-specific BSM success probability and $k$ is the number of BSM operations required along an end-to-end path. A larger k severely limits the success rate per time slot. This increased latency forces qubits to be stored for extended durations, inducing significant memory-induced decoherence. By prioritizing NCA-based paths to minimize $k$ (Eq. 2), RADAR-Q ensures that the exponential benefit of fewer BSM operations translates directly into higher end-to-end fidelity, effectively neutralizing the 'retry-induced' noise that plagues non-localized protocols.

\subsection{Conflict Resolution and Fairness}

\begin{figure}[htbp]
    \centering
    \begin{subfigure}[b]{0.49\textwidth}
        \centering
        \includegraphics[width=\textwidth]{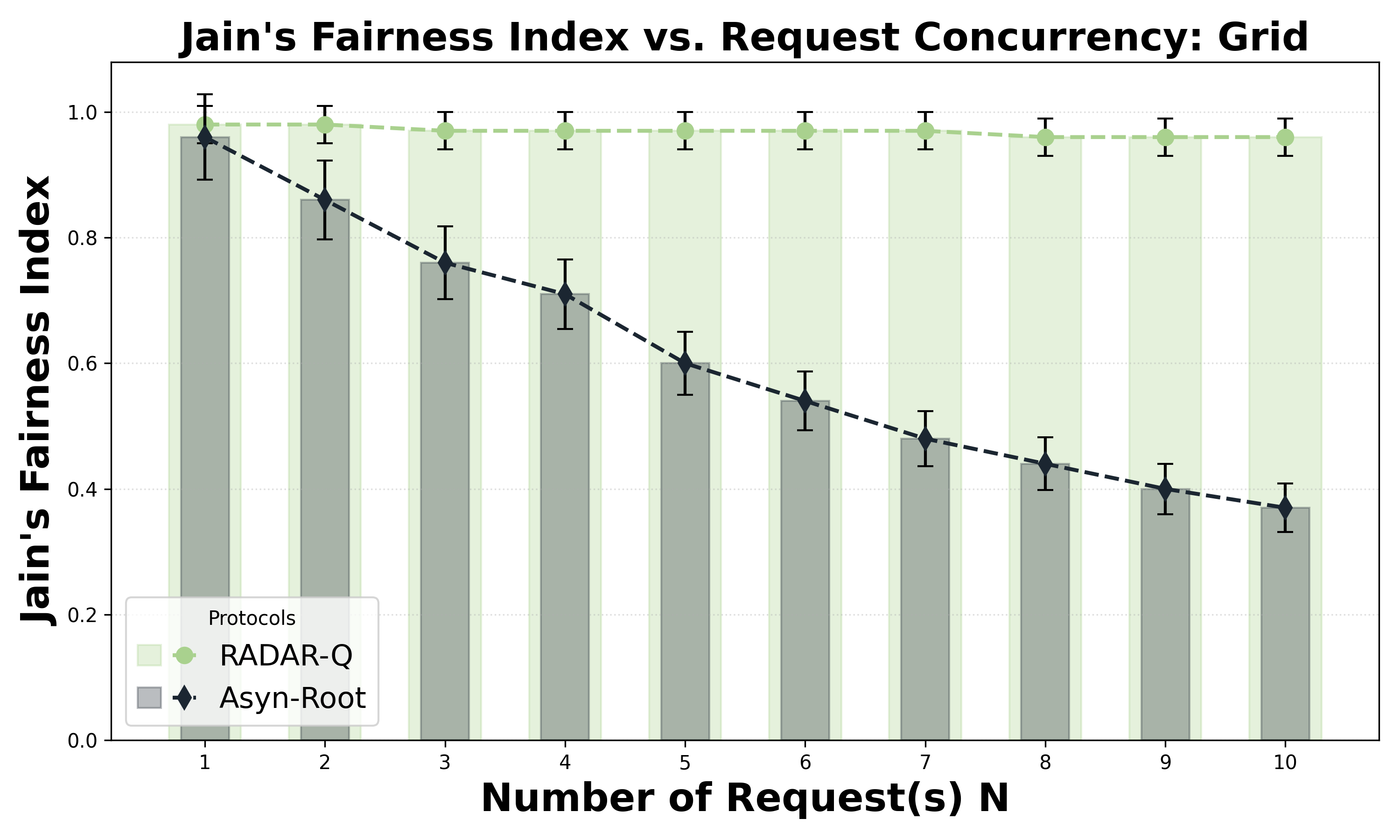}
        \caption{Grid Topology}
        \label{fig:grid_fair}
    \end{subfigure}
    \hfill
    \begin{subfigure}[b]{0.49\textwidth}
        \centering
        \includegraphics[width=\textwidth]{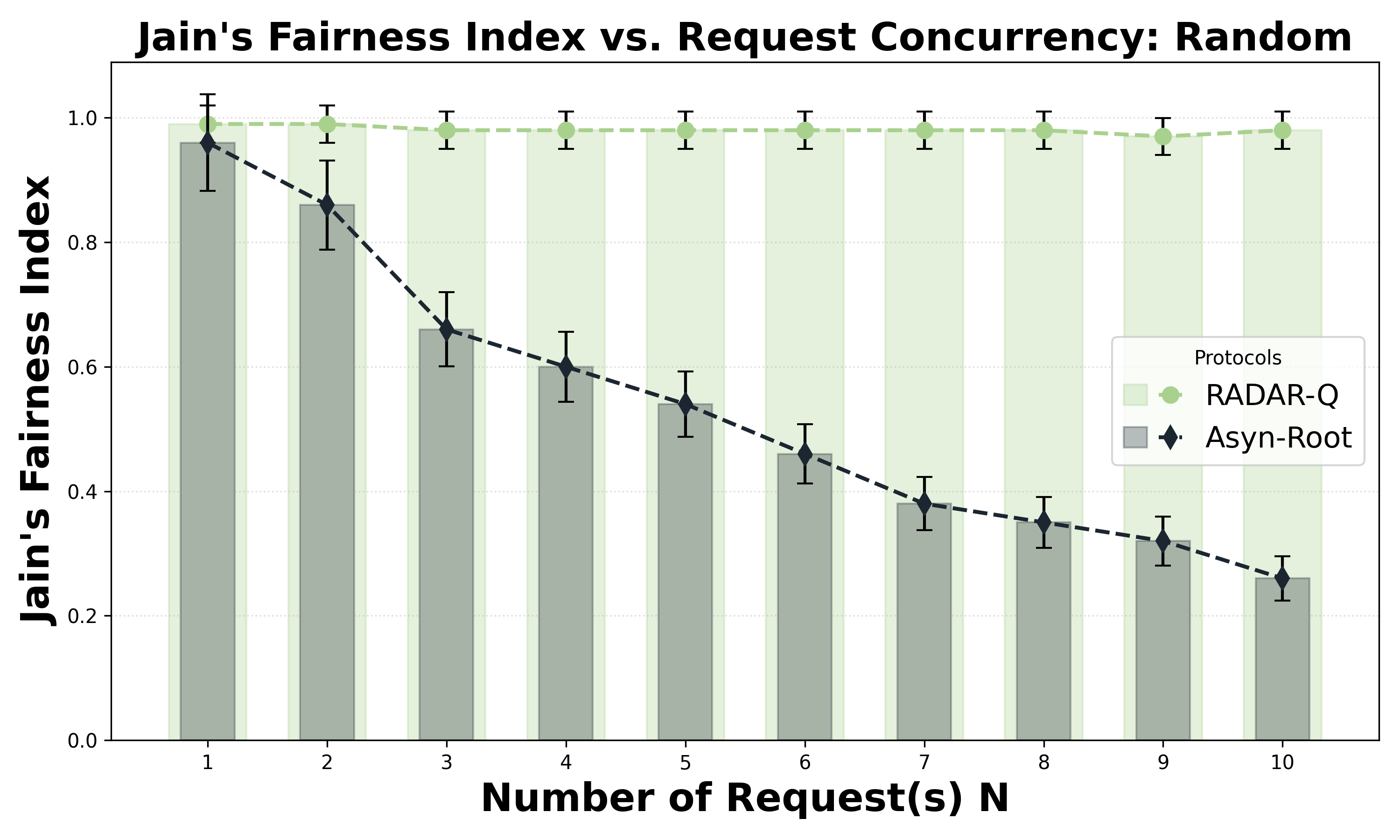}
        \caption{Random Topology}
        \label{fig:rand_fair}
    \end{subfigure}
    \caption{Jain's Fairness Index for resource allocation. RADAR-Q preserves perfect equity (Index $\approx 0.98$), preventing the structural bottlenecks that cause the 74\% collapse in Asynch-Root.}
    \label{fig:fairness_resolution}
\end{figure}

RADAR-Q also demonstrates exceptional performance in network fairness, a critical metric for multi-tenant quantum networks where users must compete for constrained entanglement resources. Figures~\ref{fig:grid_fair} and~\ref{fig:rand_fair} evaluate this using Jain's Fairness Index \cite{jain1984quantitative}, where a value of 1.0 represents perfect equity in service distribution.

RADAR-Q maintains a stable fairness profile, with the index consistently hovering between 0.96 and 0.98 across both topologies. This high level of equity demonstrates that the protocol effectively prevents resource monopolization, ensuring that aggregate throughput gains are distributed uniformly among all user pairs. In contrast, the fairness of the Asyn-Root baseline collapses as request concurrency $N$ increases, plummeting to 0.39 in the grid topology and 0.24 in the random topology at N=10. This trend confirms that in standard root-centric networks, nodes geographically closer to the root monopolize available links, leaving peripheral nodes in a state of resource starvation.

The equitable performance of RADAR-Q is driven by its proactive contention resolution. By embedding real-time link availability into the path selection metric (Eq.~\ref{eq:contention-aware_metric}), the protocol steers requests toward underutilized NCA nodes. This dynamic identification of parallel paths allows RADAR-Q to bypass the structural bottlenecks that typically root-centric asynchronous designs.

\subsection{Robustness to Coherence Time}

\begin{figure}[htbp]
    \centering
    \begin{subfigure}[b]{0.49\textwidth}
        \centering
        \includegraphics[width=\textwidth]{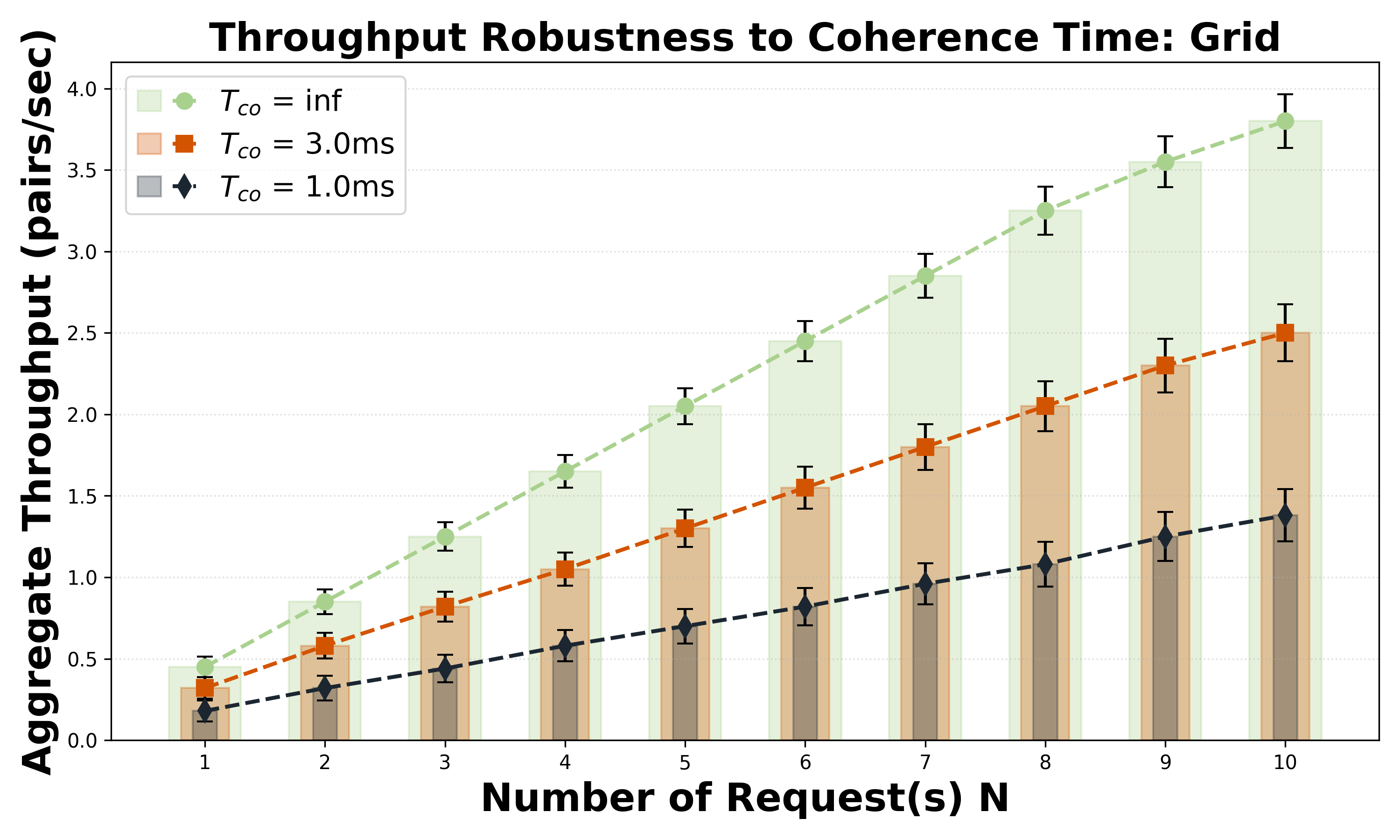}
        \caption{Grid Topology}
        \label{fig:grid_rob}
    \end{subfigure}
    \hfill
    \begin{subfigure}[b]{0.49\textwidth}
        \centering
        \includegraphics[width=\textwidth]{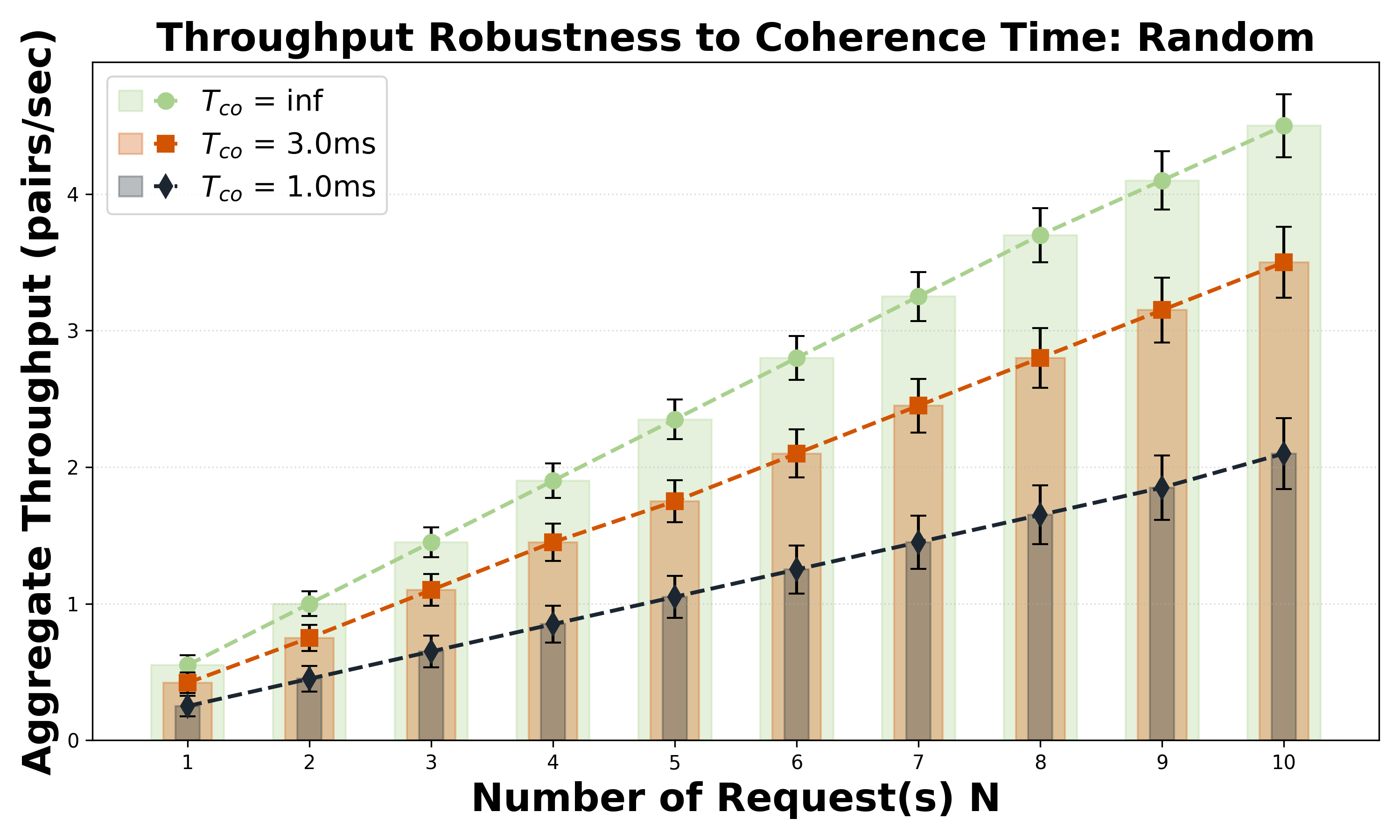}
        \caption{Random Topology}
        \label{fig:rand_rob}
    \end{subfigure}
    \caption{Throughput robustness across coherence regimes. RADAR-Q averts the scaling collapse even at $T_{co} = 1$ ms, transforming hardware volatility into a manageable penalty.}
    \label{fig:robustness_tco}
\end{figure}

Figures~\ref{fig:grid_rob} and~\ref{fig:rand_rob} evaluate RADAR-Q's performance under varying qubit coherence times ($T_{co}$), ranging from a stringent $1.0$ ms up to idealized infinity. RADAR-Q maintains linear throughput growth across all coherence regimes.

Even in the random topology at $T_{CO} = 1.0$ ms, the protocol delivers approximately 2.1~pairs/sec when $N=10$, retaining over 50\% of its idealized throughput when $T_{CO}=\infty$. This robustness is due to our NCA-centric path selection, which inherently caps the storage duration for any qubit by prioritizing short, localized entanglement segments. As a result, finite coherence time imposes only a constant multiplicative penalty on absolute throughput, rather than triggering the load-dependent failures observed in root-centric protocols. This structural decoupling of network scalability from hardware volatility makes RADAR-Q suitable for near-term NISQ-era platforms where coherence is a primary constraint.

Overall, these results establish RADAR-Q as a scalable alternative to both centralized synchronous and root-centric asynchronous designs. By making contention-aware decisions based on local state, the protocol achieves near-linear throughput growth and stable fairness without exceeding the 0.5 fidelity threshold even under the stringent coherence constraints of near-term NISQ hardware.

\section{Conclusion}
\label{sec:conclusion}

In this paper, we have demonstrated that the path to scalable multi-user quantum networking lies in local resource contention awareness rather than complex global scheduling. By embedding resource competition directly into the routing metric, RADAR-Q eliminates the structural bottlenecks inherent in traditional root-centric designs. This decentralized approach allows nodes to identify localized swapping points via the NCA, significantly reducing the BSM depth required for each session.

Our evaluation validates that this architectural shift yields substantial performance dividends: RADAR-Q delivers up to $7.6\times$ the throughput of standard asynchronous baselines while maintaining a high end-to-end fidelity of 0.76. Our results also show that while naive protocols render entanglement physically obsolete by falling below the 0.5 distillation limit, RADAR-Q's proactive path selection preserves the viability of every generated pair for downstream quantum applications. With near-perfect fairness (Jain's Index $>0.96$) and robust scalability down to 1.0~ms coherence times, RADAR-Q establishes a new benchmark for resource-efficient communication in multi-tenant quantum networks, with direct applicability to quantum data centers, distributed quantum computing, and other shared quantum infrastructures as they scale.

\section*{Acknowledgments}
\label{sec:ack}

This work was supported in part by Cisco University Research Grant \#98690499 and the Qatar Research, Development, and Innovation (QRDI) Academic Research Grant \#ARG02-0415-240191. The statements made here are solely the responsibility of the authors.   
%
%
%
\bibliographystyle{splncs04}
\bibliography{reference}
\end{document}